\documentclass[fleqn,twoside]{article}
\usepackage{espcrc2}
\usepackage{graphicx}

\newcommand{\AmS}{{\protect\the\textfont2
  A\kern-.1667em\lower.5ex\hbox{M}\kern-.125emS}}

\title{Superparamagnetic behaviour of antiferromagnetic DyPO$_{4}$ nanoparticles}

\author{T. G. Sorop\address[kol]{Kamerlingh Onnes Laboratory,
Leiden University, 2300 RA, Leiden, The Netherlands},
        M. Evangelisti\addressmark[kol]%
        \thanks{E-mail evange@phys.leidenuniv.nl.},
        M. Haase\address{Institut f\"ur Physikalische Chemie, Universit\"at Hamburg, 20146 Hamburg, Germany}
        and
        L. J. de Jongh\addressmark[kol]}

\begin{document}

\begin{abstract}
We report on the low-temperature magnetic ac-susceptibility of antiferromagnetic DyPO$_{4}$
nanoparticles with a very high surface to volume ratio. The results are interpreted in terms of
superparamagnetic relaxation of the N\'eel vector arising from a relatively large number $\approx
0.2~N_{a}$ of uncompensated spins probably existing on the surface of the nanoparticles. The
activation energy of the relaxation process is found to be $E_{a}/k_{B}=(2.6\pm 0.1)$~K within a
model taking into account the magnetic interaction between nanoparticles.

\vspace{1pc}
\end{abstract}

\maketitle

The magnetic structure of antiferromagnetic (AF) nanosized particles is usually described as an
exchange-coupled spherical bilayer composed of an inner antiferromagnetic core and an outer shell
of uncompensated spins. The exchange interaction between the two layers gives rise to an
unidirectional anisotropy which may cause superparamagnetic relaxation of the N\'eel vector due to
the uncompensated spins~\cite{superpara}. AF nanoparticles are of fundamental interest since, in
comparison with ferromagnetic nanoparticles of comparable size, they are better candidates for the
observation of magnetic quantum tunneling phenomena~\cite{mqt}.

In this brief communication we present very-low-temperature ac-susceptibility experiments on
nanoparticles of dysprosium phosphate. It is well established that in the magnetic structure of
bulk DyPO$_{4}$ the Dy moments are antiferromagnetically coupled and that they order at
3.4~K~\cite{bulk}. Moreover, strong crystal field effects (CFE) are responsible for highly
anisotropic low-lying energy levels, resulting in Ising-type behaviour. In what follows, we shall
describe the magnetic properties of small particles of DyPO$_{4}$ with a mean size of 2--3~nm and
narrow size distribution, which have been obtained by liquid-phase synthesis methods~\cite{haase}.
High crystallinity has been revealed by transmission electron microscopy and X-ray powder
diffraction measurements. Susceptibility data down to 2~K were obtained with a commercial
SQUID-based magnetometer. The ac-susceptibility in the temperature range 50~mK~$<T<6~$K was
measured with a mutual inductance technique in a dilution refrigerator. The excitation amplitude
was 10~mOe and the frequency was varied between $f=100$~Hz and 7~kHz. All data were collected on
powdered samples.

As in bulk DyPO$_{4}$, strong CFE are observed at high temperature. By lowering the temperature,
the susceptibility of the nanoparticles differs substantially from that of the bulk material; no
sign of magnetic ordering is observed at 3.4~K. Instead, a cusp in the real component
$\chi^{\prime}$ of the ac-susceptibility is found at about 1~K, and it is accompanied by a non-zero
imaginary component $\chi^{\prime\prime}$ at some lower temperature (Fig.~1). As shown in Fig.~1
both maxima in $\chi^{\prime}$ and $\chi^{\prime\prime}$ are strongly frequency dependent. Taking
into account the high degree of anisotropy in this material, the experimental data suggest
superparamagnetic blocking of the Dy moments below a blocking temperature $T_{B}$ corresponding to
the temperature at maximum absorption. Using the average value $T_{B}=0.8$~K for this range of
frequencies, the frequency shift of $T_{B}$, $\Delta T_{B}/(T_{B}\Delta\log f)$, where $\Delta
T_{B}$ is the change in $T_{B}$ for the given change in frequency ($\Delta\log f=1.8$), provides
the value of 0.18, which is comparable to that of other superparamagnets~\cite{mydosh}.

\begin{figure}[t!]
\centering\includegraphics[angle=0,width=6cm]{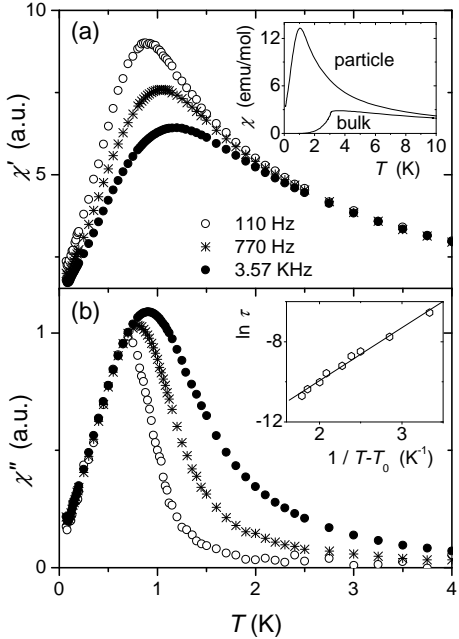} \caption{(a) $\chi^{\prime}$ vs $T$ at different
frequencies, as listed. In the inset, $\chi$ of the nanoparticles for $f=770$~Hz and $\chi$ of the
bulk material (from Ref.~\cite{bulk}) in the same units. (b) $\chi^{\prime\prime}$ vs $T$ at the
same frequencies. In the inset, the Vogel-Fulcher fit to the relaxation time $\tau$.}
\end{figure}

At temperatures below $T_{B}$, $\chi^{\prime\prime}$ does not depend on the frequency anymore
(Fig.~1). Such a behaviour is known to be related to interparticle interactions (e.g. of dipolar
origin), slowing down the relaxation at low temperature~\cite{dipole}. Taking into account this,
the frequency dependence of $T_{B}$ should be fitted using the Vogel-Fulcher law
$\tau=\tau_{0}\exp[E_{a}/k_{B}(T-T_{0})]$, where $\tau=1/2\pi f$, $E_{a}$ is the activation energy,
and $T_{0}$ represents a measure of a static interaction field due to the surrounding particles.
The fit is shown in the bottom inset of Fig.~1, and provides the following values: $\tau_{0}=(2\pm
1)\times 10^{-7}$~s, $E_{a}/k_{B}=(2.6\pm 0.1)$~K and $T_{0}=(0.40\pm 0.05)$~K. As in other
magnetically interacting nanoparticle systems, $T_{0}\ll E_{a}/k_{B}$.

It is worthwhile to give an estimate of the number of unpaired Dy spins that are responsible for
the net magnetic moment of the particles. To do so we plot in the top inset of Fig.~1 the
susceptibility of DyPO$_{4}$ nanoparticles together with the parallel susceptibility measured from
a single-crystal of the bulk equivalent material~\cite{bulk}. The susceptibility of the
nanoparticles is here properly scaled to get the same values as for the bulk material at high
temperatures. The striking feature is the large difference between the two signals in the
low-temperature range which we suppose to arise from a lack of compensation of the
antiferromagnetic sublattices in the nanoparticles. We consider the $T<20~$K range, where only the
lowest electronic doublet is contributing to the magnetic signals, and fit the susceptibilities
well above the anomalies with a Curie-Weiss law. Comparing the obtained Curie constants, we roughly
estimate the number of unpaired spins per particle to be about 20\%. Considering that the mean
particle diameter is 2.5~nm~\cite{haase}, which corresponds to about 130 Dy spins per particle, we
roughly have about 30 spins antiferromagnetically arranged in the inner core, and about 75
compensated and 25 uncompensated spins on the surface. The presence of uncompensated spins is thus
most probably caused by missing magnetic neighbours at the surface.

This work was part of the research program of the Stichting voor FOM.

\end{document}